\documentclass[10pt,twocolumn,showpacs,aps,floatfix]{revtex4}
\usepackage{epsfig}
\usepackage{dcolumn}
\usepackage{bm}

\newcommand{\la}{\langle}
\newcommand{\ra}{\rangle}

\begin{document}



\title{Excited states of the helium-antihydrogen system.}

\author{Vasily Sharipov$ ^{1,3}$, Leonti N. Labzowsky$ ^{1,2,3}$, and G\"unter Plunien$^{3}$}

\affiliation{$^{1}$Institute of Physics, St. Petersburg State University, 198904, Uljanovskaya 1,
Petrodvorets, St. Petersburg, Russia}
\affiliation{$^{2}$Petersburg Nuclear Physics Institute, 188350,
Gatchina, St. Petersburg, Russia}
\affiliation{$^{3}$Institut f\"ur Theoretische Physik, 
Technische Universit\"at Dresden, Mommsenstrasse 13,
D-01062, Dresden, Germany}



\begin{abstract}
Potential energy curves for excited leptonic states of the helium-antihydrogen system are calculated within Ritz' variational approach. An explicitly correlated ansatz for the leptonic wave function
is employed describing accurately the motion of the leptons (two electrons and positron) in the field of the helium nucleus and of the antiproton with arbitrary orbital angular momentum projection $\Lambda$ onto the internuclear axis. Results for $\Lambda$=0, 1 and 30 are presented. For quasibound states with large values of $\Lambda$ and rotational quantum numbers $J>\Lambda$ no annihilation and rearrangement decay channels occur, i. e. they are metastable.
\end{abstract}



\maketitle


Two groups at CERN \cite{Amoretti,Gabrielse} reported recently on the production of antihydrogen  
(\=H) atoms, which under the present experimental conditions  were produced in Rydberg states.
Future experiments with \=H are aiming for investigations of their spectroscopical 
properties inside of a trap. 
The experimental progress achieved  has stimulated theoretical investigations of atom-antiatom systems. Corresponding calculations are usually based on 
variational methods utilizing the Born-Oppenheimer approximation.
Several works devoted to the H-\=H system were considering both scattering phenomena (see, e.g. Refs. \cite{Armour,Zygelman}) and the formation of quasibound states \cite{Zygelman,Labzowsky}.  
The potential energy curve of the H\=H system was found to be a monotonic function
of the  internuclear distance $R$. 
There exists a so-called critical internuclear distance $R^{}_{\rm c}$, where the electron-positron (e$^{-}$e$^{+}$) pair becomes unbound to the nuclei (proton and antiproton, p$^{+}$p$^{-}$). Close 
to $R^{}_{\rm c}$ the Born-Oppenheimer approximation breaks down. 
The optical potential method developed in \cite{Zygelman} is aiming for an adequate description of the positronium (Ps atom) ejection within the framework of the adiabatic picture. 
The interaction between the He and \=H atom both being in their ground states seems also well understood \cite{Strasburger1,Jonsell3,Strasburger2}. A small potential wall has been discovered \cite{Strasburger3}, which allows for  
a few quasibound states with lifetimes of about  $10^{-12}-10^{-10}$ sec 
\cite{Jonsell3}. The Born-Oppenheimer approach also applies for describing the ground state of the He\=H system. 
For investigations of antihydrogen-atom collisions non-adiabatic methods have been developed as well. 
Cross sections for scattering of \=H on the hydrogen \cite{Sinha1}, helium \cite{Sinha2} and alkali-metal atoms \cite{Sinha3} have been calculated employing the atomic orbital expansion technique.

Since the \=H atoms are produced in Rydberg states, the atom-\=H system will be most likely formed in states with high values of orbital angular momentum projection $\Lambda$ onto the internuclear axis. From a pragmatic point of view it is important to develop an adequate description for such states. This was already achieved for the H\=H system in \cite{Sharipov1,Sharipov2} by utilizing 
explicitly correlated wave functions,  which describe accurately the motion of leptons with arbitrary orbital angular momentum projection. 
In the present Letter we proceed in a similar way and extend our approach 
to describe the interaction between a He atom in its ground (singlet) state and an \=H$^{*}$ atom 
in an excited state. Atomic units will be used throughout.

Application of the adiabatic approach  to the He\=H system leads to the Schr\"odinger equation 
\begin{equation}
\hat{H}^{\rm lep}\Psi^{\rm lep}_{\Lambda}({\bf x},{\bf R})=V^{\rm lep}_{\Lambda}(R)\Psi^{\rm lep}_{\Lambda}({\bf x},{\bf R})  \label{Sel}
\end{equation}
for the wave function $\Psi^{\rm lep}_{\Lambda}$ describing the motion of the leptons  in the field 
of the He nucleus ($\alpha$-particle) and of the antiproton 
Here ${\bf x}=({\bf r}^{}_{1},{\bf r}^{}_{2},{\bf r}^{}_{3})$ denotes the position vectors
of the leptons and ${\bf R} = {\bf R}^{}_{\alpha} - {\bf R}^{}_{{\rm p}^{-}}$ defines the
internuclear distance $R=|{\bf R}|$ between the nuclei located at ${\bf R}^{}_{\alpha}$ and  
${\bf R}^{}_{{\rm p}^{-}}$ with respect to the center-of-mass frame.
The leptonic Hamiltonian reads  
\begin{eqnarray}
\hat{H}^{\rm lep} &=&-\frac{1}{2}\sum_{i=1}^3\triangle^{}_{i}+\sum_{i=1}^3\frac{2e^{}_{i}}{|{\bf r}^{}_{i}-{\bf R}^{}_{\alpha}|} \nonumber \\
&&-\sum_{i=1}^3\frac{e^{}_{i}}{|{\bf r}^{}_{i}-{\bf R}^{}_{{\rm p}^{-}}|}+\sum_{i\neq j}^3\frac{e^{}_{i}e^{}_{j}}{|{\bf r}^{}_{i}-{\bf r}^{}_{j}|} ,
\end{eqnarray}
where  $e^{}_{i,j}=\mp 1$ refers to the charge of e$^{-}$ and e$^{+}$, respectively.
The He\=H interaction energy is expressed in terms of the  leptonic potential $E^{}_{\Lambda}(R)=V^{\rm lep}_{\Lambda}(R)-2/R$.

Axial symmetry implies the eigenvalue equation 
\begin{equation}
\hat{L}^{}_{\bf R}\Psi^{\rm lep}_{\Lambda}({\bf x},{\bf R})=\Lambda\Psi^{\rm lep}_{\Lambda}({\bf x},{\bf R}) \label{Mp}
\end{equation}
for the component $\hat{L}^{}_{\bf R}$ of the leptonic orbital angular momentum operator along the internuclear axis 
with eigenvalue (orbital angular momentum projection) $\Lambda$. 
If the internuclear distance tends to infinity the He and \=H atoms no longer interact. Thus, in the limit $R\to\infty$ the leptonic wave function equals the product
\begin{equation}
\Psi^{\rm lep}_{\Lambda}({\rm He}\bar{\rm H})=\Psi^{}_{\rm gs}({\rm He})\Psi^{}_{\rm es}(\bar{\rm H})  \label{as}
\end{equation}
between the wave functions $\Psi^{}_{\rm gs}({\rm He})$ and $\Psi^{}_{\rm es}(\bar{\rm H})$ describing the He atom in the ground state (gs) and the \=H$^{*}$ atom in an excited state (es), respectively. In the case of large internuclear distances 
the leptonic orbital angular momentum projection $\Lambda$ of the He\=H$^{*}$ system is mainly 
carried by the e$^{+}$.

For solving the Schr\"odinger equation (\ref{Sel}) an explicitly correlated ansatz for the leptonic wave function is taken 
\begin{eqnarray}
\Psi^{\rm lep}_{\Lambda}({\bf x},{\bf R})&=&\sum_{k=1}^N C^{}_{k}\psi^{k}_{\Lambda}({\bf x},{\bf R}),\label{dec} \\
\psi^{k}_{\Lambda}({\bf x},{\bf R})&=&\hat{P}\exp{\left(-\sum_{i=1}^3 a^{k}_{i}({\bf r}^{}_{i}-{\bf R}^{k}_{i})^{2}\right)}\times \nonumber \\
&\times&\exp{\left(-\sum_{i\neq j}^3 b^{k}_{ij}({\bf
r}^{}_{i}-{\bf r}^{}_{j})^{2}\right)}\cdot\eta^{k}_{\Lambda}({\bf
x},{\bf R}),\label{ecg}\\
\eta^{k}_{\Lambda}({\bf x},{\bf R})&=&|{\bf
v}^{}_{k}|^{\Lambda}Y^{}_{\Lambda\Lambda}(\hat{\bf v}^{}_{k}),\label{ang}
\end{eqnarray}
where we defined the vectors ${\bf v}_k=\sum_{i=1}^3 u^{k}_{i}{\bf r}^{}_{i}$. The coefficients $C^{}_{k}$ are linear and $b^{k}_{ij}$, $a^{k}_{i}$, ${\bf R}^{k}_{i}$ and $u^{k}_{i}$ are nonlinear variational parameters, while the operator $\hat{P}$ ensures proper symmetry. 
In our calculation a set of $N=300$ basis functions 
$\{\psi^{k}_{\Lambda}\}_{k=1}^N$
was adopted. Each function $\psi^{k}_{\Lambda}$ appears as product of Explicitly Correlated Gaussians (ECGs) together with an angular part $\eta^{k}_{\Lambda}$. 
The ECGs alow for an adequate description of the lepton in the field of the nuclei.
The angular part $\eta^{k}_{\Lambda}$ involving a usual spherical harmonic 
$Y^{}_{\Lambda\Lambda}$ ensures the proper angular symmetry of 
each basis function  $\psi^{k}_{\Lambda}$ according to equation (\ref{Mp}). 

For any fixed internuclear distance each nonlinear parameter was optimized employing the golden section method. Though there are more refined optimization methods, e. g. \cite{Cencek}, we prefer to use this very simple but reliable approach. For few-particle systems such as H\=H or He\=H quasimolecules it does not look much more cumbersome than other more sophisticated approaches. The set of linear variational parameters $\{C^{}_{k}\}^{N}_{k=1}$ was obtained by solving the generalized eigenvalue problem.

\begin{table}
\begin{tabular}{p{0.9cm}|ccc}
\multicolumn{4}{p{7cm}}{Table 1: The interaction energy $E^{}_{\Lambda}(R)$ as a function of the internuclear distance $R$ is calculated for leptonic orbital angular momentum projections $\Lambda$=0, 1, 30 (in atomic units).}\\[5pt]
\hline
$R$ & $E^{}_{0}(R)$ & $E^{}_{1}(R)$ & $E^{}_{30}(R)$ \\
\hline
0.2 & -10.8411 & -10.7964 & -10.6110 \\
0.4 & -5.98124 & -5.91664 & -5.72835 \\
0.6 & -4.51584 & -4.40717 & -4.21115 \\
0.8 & -3.91018 & -3.73842 & -3.58723 \\
1.2 & -3.50499 & -3.25801 & -3.12888 \\
1.4 & -3.44579 & -3.16847 & -3.04498 \\
1.6 & -3.42218 & -3.11499 & -2.99453 \\
1.8 & -3.41023 & -3.08854 & -2.96606 \\
2.0 & -3.40524 & -3.06927 & -2.94541 \\
2.2 & -3.40344 & -3.05695 & -2.92954 \\
2.5 & -3.40310 & -3.04585 & -2.91894 \\
2.7 & -3.40334 & -3.04132 & -2.91609 \\
2.9 & -3.40363 & -3.03814 & -2.91413 \\
3.1 & -3.40388 & -3.03590 & -2.91179 \\
3.3 & -3.40402 & -3.03428 & -2.90986 \\
3.5 & -3.40411 & -3.03308 & -2.90836 \\
3.8 & -3.40415 & -3.03182 & -2.90702 \\
4.0 & -3.40410 & -3.03123 & -2.90698 \\
5.0 & -3.40390 & -3.02969 & -2.90532 \\
7.0 & -3.403729 & -3.028904 & -2.904478 \\
10.0 & -3.403706 & -3.028712 & -2.904259 \\
12.0 & -3.4037047 & -3.0286969 & -2.9042220 \\
15.0 & -3.4037042 & -3.0286939 & -2.9041912 \\
20.0 & -3.4037040 & -3.0286926 & -2.9039813 \\[5pt]
\hline
\end{tabular}
\end{table}

\begin{figure}
\epsfig{file=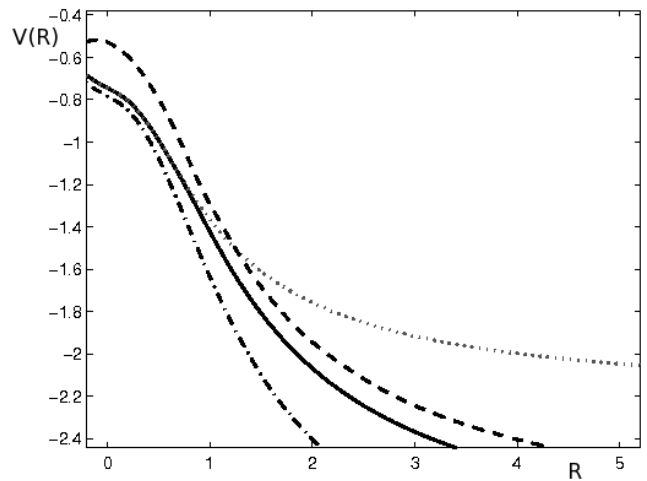, bbllx=60pt, bblly=0pt, bburx=660pt,
bbury=600pt, height=7.5cm} \caption{The leptonic potentials for
the ground state ($\Lambda =0$) and for excited states ($\Lambda = 1, 30$), $V^{\rm lep}_{0}(R)$ (dash-dotted line), $V^{\rm lep}_{1}(R)$ (solid line) and $V^{\rm lep}_{30}(R)$ (dashed line) are plotted versus the internuclear distance $R$ (in atomic units). 
The potential for the lowest state of the He$^{+}$p$^{-}$+unbound Ps system $V^{}_{\rm Ps}(R)$ (dotted line) is also plotted. The potential $V^{\rm lep}_{1}(R)$ approaches $V^{}_{\rm Ps}(R)$ near the critical internuclear distance $R^{}_{\rm c}=0.7a^{}_{0}$.}\label{helep}
\end{figure}

Results for the He\=H interaction energies $E^{}_{\Lambda}$ for $\Lambda$=0, 1 and 30 are presented in Table~1. The behavior of the potential curve for the ground state of the He\=H system ($\Lambda$=0) was first examined in \cite{Strasburger3}. Our results for $\Lambda$=0 are in a good agreement with previous investigations \cite{Strasburger3} and achieved with half the number of 
basis functions. Furthermore, we shall focus on the excited levels of the He\=H system.

The leptonic potentials $V^{\rm lep}_{\Lambda}$ for He\=H with $\Lambda$=0, 1 and 30 are depicted in Fig.~\ref{helep} together with the potential $V^{}_{\rm Ps}$ for a 
compound He$^{+}$p$^{-}$ + unbound Ps system in its lowest state. We assume, that  
for a state with a certain value of $\Lambda$ the 
Ps ejection occurs, if the potential energy curve $V^{\rm lep}_{\Lambda}$ coincides 
with the curve $V^{}_{\rm Ps}$ at distances $R\le R^{}_{\rm c}(\Lambda)$. In principle, the crossing of the two curves does not immediately imply the ejection of a Ps atom. With respect to the nuclear center-of-mass system the energy $E^{}_{\rm Ps}=E^{\rm bind}_{\rm Ps}+E^{\rm rot}_{\rm Ps}$
of the e$^{-}$e$^{+}$ pair consists of two parts: 
the binding energy $E^{\rm bind}_{\rm Ps}$ of the ground state of the Ps atom and 
the rotational energy $E^{\rm rot}_{\rm Ps}$ of the Ps atom with respect to the internuclear axis. 
For values $\Lambda > 0$ but small, the energy $E^{\rm rot}_{\rm Ps}$ is negligible compared to $|E^{\rm bind}_{\rm Ps}|$. Conversely, for $\Lambda\gg 1$ the contribution $E^{\rm rot}_{\rm Ps}$ is essential as it was shown in \cite{Sharipov1,Sharipov2} for the H\=H$^{*}$ system. The rotational part of the energy  $E^{\rm rot}_{\rm Ps}$ is not taken into account in the potential 
for the He$^{+}$p$^{-}$+unbound Ps system, since it  is inversely proportional to the 
square of the distance between the Ps atom and the He$^{+}$p$^{-}$ compound.
Thus, the crossing of the potential curves $V^{}_{\rm Ps}$ and $V^{\rm lep}_{\Lambda}$ does not 
mean that the energies of the two systems (He\=H$^{*}$ and He$^{+}$p$^{-}$+Ps) become equal. 
For the states with large values of $\Lambda$ this takes place only, when the internuclear distance $R$ becomes sufficiently small and the Ps atom appears to be far enough from the He$^{+}$p$^{-}$ 
compound, so that the rotational energy becomes negligible. 
Consequently, the curves for the He\=H$^{*}$ and the combined He$^{+}$p$^{-}$+Ps systems 
coincide for all values of $R\le R^{}_{\rm c}$. 

According to Fig.~\ref{helep} the properties of the states of the He\=H system with the $\Lambda$ values under consideration ($\Lambda$=0, 1, 30) differ strongly. The He\=H potential energy curve  $V^{\rm lep}_{0}$ does not cross the one for the He$^{+}$p$^{-}$+unbound Ps system. 
Therefore, the light particles appear to be bound in the He\=H system over the entire range of internuclear distances $R$. 
For $R\to 0$ the function $V^{\rm lep}_{0}$ approaches the binding energy of the positronium hydride $E^{}_{\rm HPs}=-0.7891967$ \cite{Yan}. This reveals that at small internuclear distances the He\=H quasimolecule transforms into the He$^{+}$p$^{-}$ system plus the Ps atom weakly attached to it. The He\=H$^{*}$ states with nonzero but small $\Lambda$ (see e.g. $\Lambda$=1) exhibit properties similar to those of the H\=H$^{*}$ system \cite{Sharipov1,Sharipov2}. There exists a critical internuclear distance $R^{}_{\rm c}(\Lambda=1)=0.7a^{}_{0}$ at which the wave function $\Psi^{\rm lep}_{\Lambda}$ transforms from the wave function of a bound He\=H$^{*}$ system into that of an unbound Ps atom in the field of the He$^{+}$p$^{-}$. 
The latter wave function then contains a plane wave factor describing the center-of-mass motion of the Ps atom. Thus, the adiabatic correction to the leptonic potential diverges near $R^{}_{\rm c}$ indicating the breakdown of the Born-Oppenheimer approximation in the vicinity of the 
critical distance. 
Despite the fact that the ECGs cannot properly reproduce a plane wave, we keep the ansatz Eqs.~(\ref{dec},\ref{ecg},\ref{ang}) even for internuclear distances $R\le R^{}_{\rm c}$. 
However, the basis set (\ref{ecg}) provides the correct value for the energy of the lowest continuum state of the He$^{+}$p$^{-}$+unbound Ps system (i. e. with zero relative velocity of Ps atom with respect to He$^{+}$p$^{-}$) as well as for the matrix element involving a spatial delta-function (see below). Properties of the states with large values of $\Lambda$ have already been elucidated for the H\=H system in \cite{Sharipov1,Sharipov2}. 
As  implied by Eq.~(\ref{as}), for large internuclear distances the He\=H$^{*}$ system 
can be envisaged as a He atom plus a p$^{-}$ and e$^{+}$ weakly attached to them. 
As the internuclear distance $R$ decreases the system changes slightly in the following sense: 
The orbital angular momentum can still be mainly attributed to the e$^{+}$. Actually, the potential curves for the He\=H$^{*}$ and Hep$^{-}$ systems behave similarly \cite{Gibbs}. At $R=0$ the potential $V^{\rm lep}_{30}$ approaches the value of the binding energy of an H$^{-}$ ion plus a small contribution from a weakly bound e$^{+}$ (see Fig.~\ref{helep}).   
  H\=H$^{*}$ and He\=H$^{*}$ systems in states with high $\Lambda$, respectively, 
differ mainly in their behavior at small internuclear distances. When the p$^{-}$ approaches the p$^{+}$ in H\=H$^{*}$ system, the e$^{-}$ becomes loosely bound and forms a Ps atom together with  the e$^{+}$. In case of the He\=H$^{*}$ system, the $\alpha$-particle tends to keep both  e$^{-}$s due to its larger electric charge and the Ps atom is not ejected. According to Fig.~\ref{helep}, the leptonic potentials for the He\=H$^{*}$ with $\Lambda =30$ and that of the He$^{+}$p$^{-}$+unbound Ps system cross each other. However, as explained above, this does not imply the ejection of the Ps atom for the He\=H$^{*}$ system in the state with $\Lambda =30$.

\begin{figure}
\epsfig{file=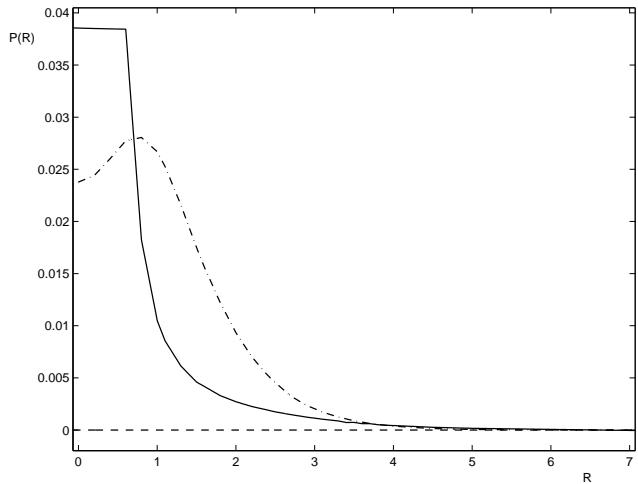, bbllx=60pt, bblly=180pt, bburx=600pt,
bbury=600pt, height=7cm} \caption{The coalescence probability distributions $P^{}_{0}(R)$ (dash-dotted line), $P^{}_{1}(R)$ (solid line) and $P^{}_{30}(R)$ (dashed line) are plotted as a function of the internuclear distance $R$ (in atomic units).}\label{heann}
\end{figure}

The results discussed above can be confirmed by evaluating the coalescence probability distribution $P^{2\gamma}_{\Lambda}$ \cite{Strasburger2} as a function of $R$, which arises 
in calculations of the two-photon leptonic annihilation rate. 
According to \cite{Ryzhikh} the general expression for $P^{2\gamma}_{\Lambda}$ is
\begin{equation}
P^{2\gamma}_{\Lambda}(R)=\la\Psi^{\rm lep}_{\Lambda}|\sum_{i=1}^{n^{}_{e^{-}}}\sum_{j=1}^{n^{}_{e^{+}}}\delta({\bf r}^{}_{i}-{\bf r}^{}_{j})(1-\hat{S}^{2}_{i,j})|\Psi^{\rm lep}_{\Lambda}\ra,
\end{equation}
where the spin operator $1-\hat{S}^{2}_{i,j}$ ensures that the $2\gamma$-annihilation can only take place between e$^{-}$ and e$^{+}$ being in the singlet state. The annihilation rate of a bound state of a particle and antiparticle is proportional to the matrix element of the spatial delta function. The coalescence probability distributions for the He\=H system with $\Lambda$=0, 1 and 30 are depicted in Fig.~\ref{heann}. If the Ps atom is ejected, the function $P^{2\gamma}_{\Lambda}$ should approach the value of the coalescence probability for the ground state of the positronium $|\psi^{}_{\rm Ps}(0)|^{2}=1/8\pi\approx 0.038$. 

In the highly excited He\=H system with $\Lambda$=30 the density of the very weakly bound e$^{+}$ 
is thinly distributed over the entire range of internuclear distances. 
Thus, the probability $P^{2\gamma}_{30}$ for the  e$^{+}$ to coalesce with the e$^{-}$ 
is negligible for all values of $R$ (see Fig.~\ref{heann}). 
On the other hand, for  $\Lambda$=30 the Ps atom (bound or unbound) is never formed within this adiabatic picture. Consequently, the leptonic annihilation decay rates and the Ps ejection decay rates are negligible for the He\=H system in the states with high values of $\Lambda$. In particular, for the quasibound states of He\=H$^{*}$ with $\Lambda$=30 and a rotational quantum number $J>\Lambda$ the angular momentum barrier prevents the $\alpha$-particle and p$^{-}$ to coalesce. Hence, the nuclear annihilation decay rates vanish for such quasibound states. Since there are no annihilation and Ps ejection decay channels, the He\=H$^{*}$ quasimolecule becomes metastable
in these states. The decay of such states with high values of $\Lambda$ and $J>\Lambda$ can occur only via a radiative cascade into a final state, where the annihilation process is probable. 
The lifetime of such  metastable states of the He\=H$^{*}$ quasimolecule are expected to be of order $10^{-6}$ s as for Hep$^{-}$ atomcules \cite{Iwasaki,Hayano}; as mentioned above the properties of these systems are similar.

Under the experimental conditions reported in \cite{Amoretti,Gabrielse} 
the \=H atoms are produced inside of traps with very high magnetic fields (up to $5$ T). 
Atomic levels with different angular quantum numbers $l$ will be fully admixed and remain so,
when the quasimolecules are formed. If the molecular axis is oriented parallel to the magnetic field, its presence will not lead to any qualitative difference compared to the case of zero field. According to the ``guiding center atoms'' picture of the three-body interaction in a plasma 
developed in \cite{Glinsky,Robicheaux}, the formation of H\=H and He\=H quasimolecules should be most probable with this orientation. Moreover, the magnetic field influences more strongly the quasimolecular (rotational) levels rather than the behavior of the potential curves; the latter being
the major subject of the present Letter.

Summarizing, we can state that the method employing an explicitly correlated ansatz as developed recently for describing Rydberg states of the H\=H system, can be successfully applied for calculations of the He\=H system. 
Accurate potential energy curves for the quasimolecular states with $\Lambda$=0, 1 and 30 are obtained. The results for the ground state of the He\=H system ($\Lambda$=0) are in agreement with the known ones. The potential energy curves and coalescence probability distributions obtained 
can be used for evaluating of the spectrum of the He\=H quasimolecule, decay rates and cross sections for various processes. The prediction of metastable states in the He\=H system (e.g. $\Lambda$=30 and $J>\Lambda$) is the most important result of the present Letter. This leads to the possibility to deposit a big amount of energy at atomic scales ($\sim 1$ GeV per molecule) over a relatively long time period ($\sim 10^{-6}$ s). Assuming typical atom velocities ($\sim 10^{7}$ cm/s) this implies a possible transfer of this energy over macroscopic distances. 
The problem of coexistence of matter and antimatter is of fundamental interest not only for laboratory studies but also for cosmology. The enormous enhancement of the  lifetime of 
Rydberg states of the He\=H$^{*}$ quasimolecule compared to H\=H$^{*}$ 
(about $10^{8}$) may have most important consequences.

The authors acknowledge financial support from INTAS-GSI grant Nr. 06-1000012-8881 and DFG. 
V.~S. and L.~L. are grateful to the TU Dresden for hospitality. The work of V.~S. and L.~L. was also supported by RFBR grant Nr 05-02-17483 and by non-profit foundation "Dynasty". G.~P. also acknowledges financial support from BMBF, DAAD and GSI.



\end{document}